\documentclass[12pt,english,longbibliography,nofootinbib,superscriptaddress,12pt,sort&compress,showkeys]{article}
\usepackage{ae,aecompl}
\usepackage[T1]{fontenc}
\usepackage[latin9]{inputenc}
\usepackage[active]{srcltx}
\usepackage{amsmath}
\usepackage{graphicx}
\usepackage[numbers]{natbib}

\makeatletter

\newcommand{\lyxaddress}[1]{
	\par {\raggedright #1
	\vspace{1.4em}
	\noindent\par}
}

\usepackage{aecompl}\usepackage{epstopdf}

\usepackage[raggedrightboxes]{ragged2e}

\usepackage{babel}

\makeatother

\usepackage{babel}
\begin{document}
\title{Atomistic modeling of diffusion processes at Al(Si)/Si(111) interphase
boundaries obtained by vapor deposition}
\author{Yang Li, Raj K. Koju and Yuri Mishin}
\maketitle

\lyxaddress{Department of Physics and Astronomy, MSN 3F3, George Mason University,
Fairfax, Virginia 22030, USA}
\begin{abstract}
\noindent Molecular dynamics and parallel-replica dynamics simulations
are applied to investigate the atomic structures and diffusion processes
at ${\text{Al}\{111\}}\parallel{\text{Si}}\{111\}$ interphase boundaries
constructed by simulated vapor deposition of Al(Si) alloy on Si(111)
substrates. Different orientation relationships and interface structures
are obtained for different pre-deposition Si (111) surface reconstructions.
Diffusion of both Al and Si atoms at the interfaces is calculated
and compared with diffusion along grain boundaries, triple junctions,
contact lines, and threading dislocations in the Al-Si system. It
is found that ${\text{Al}\{111\}}\parallel{\text{Si}}\{111\}$ interphase
boundaries exhibit the lowest diffusivity among these structures and
are closest to the lattice diffusivity. In most cases (except for
the Si substrate), Si atoms are more mobile than Al atoms. The diffusion
processes are typically mediated by Al vacancies and Si interstitial
atoms migrating by either direct or indirect interstitial mechanisms. 
\end{abstract}

\maketitle

\section{Introduction\label{sec:Intro}}

Diffusion along metal-nonmetal interfaces plays an important role
in many processes, such as phase transformations, diffusional creep
in composite materials, and microstructure development. Despite the
practical significance, the fundamental understanding of diffusion
in metal-nonmetal interfaces remains highly incomplete. Experimental
investigations of interphase boundary diffusion are extremely challenging
and reliable information is scarce. There have been very few measurements
for metal-nonmetal systems \citep{Straumal:1984aa,Kosinova:2015aa,Kumar:2018aa,Barda:2020aa},
most of which are indirect. In particular, there have been no direct
or indirect diffusion measurements for Al-Si interphase boundaries.
Meanwhile, this system presents significant interest in the context
of Al-Si composite alloys and as a model system to understand the
phenomenon of interfacial creep in metal-matrix composites \citep{Funn:1998aa,Dutta:2000aa,Peterson:2003aa,Peterson:2002aa,Peterson:2004aa}. 

Previous research \citep{Al-Si-IPB,Chesser:2024aa} applied atomistic
simulation methods to compute Al-Si interface diffusion coefficients
and understand the underlying diffusion mechanisms. A critical step
in such studies is to construct interfaces that remain structurally
stable during high-temperature anneals required for diffusion calculations.
In \citep{Al-Si-IPB,Chesser:2024aa}, the interfaces were constructed
by either bonding Si and Al single crystals with pre-determined crystallographic
orientations, or by solidifying a liquid Al(Si) alloy on a Si substrate.
Some of the stable orientation relationships discovered in \citep{Al-Si-IPB,Chesser:2024aa}
had been previously observed in epitaxy experiments. Unfortunately,
diffusion of Al and Si in the stable Al-Si interfaces was found to
be too slow to quantify by conventional molecular dynamics (MD) simulations.
Interface disconnections accelerated the interface diffusion, enabling
calculations of diffusion coefficients. These turned out to be much
smaller compared to grain boundary (GB) diffusion in Al. However,
the intrinsic interface diffusivity in the absence of extrinsic defects
remained beyond the reach of the MD methodology.

In this paper, we report on new atomistic simulations addressing the
above limitation and providing a clearer picture of where the Al-Si
interfaces belong in the hierarchy of diffusion pathways in materials.
We focus on ${\text{Al(Si)}\{111\}}\parallel{\text{Si}}\{111\}$ interfaces
created by simulated vapor deposition. We show that the orientation
relationship and interface structure depend on the pre-deposition
reconstruction of the Si(111) surface. In all cases studied, the interface
remains stable during the subsequent anneals at all temperatures up
to the eutectic point. In contrast to previous work \citep{Al-Si-IPB,Chesser:2024aa},
we combine conventional MD with the parallel replica dynamics method,
which has allowed us to evaluate the \emph{intrinsic} interface diffusivity
and gain insights into the interface diffusion mechanisms. Furthermore,
we have been able to calculate the diffusion coefficients in GBs,
triple junctions, contact lines, and dislocations in the \emph{same}
system. This has enabled us to compare the Al-Si interface diffusivities
with those in other short-circuit diffusion paths as well as the lattice
of the Al(Si) alloy. The results indicate that the interphase diffusivity,
at least in the ${\text{Al}\{111\}}\parallel{\text{Si}}\{111\}$ system,
is extremely low and closest to the lattice diffusivity in the ranking
of different diffusion pathways.

\section{Methodology}

\subsection{Vapor deposition simulations}

The Al/Si interphase boundaries (IPBs) were constructed through MD
simulations mimicking epitaxy experiments. The MD simulations were
conducted using the Large-scale Atomic/Molecular Massively Parallel
Simulator (LAMMPS) \citep{Plimpton95}. The interatomic potential
developed by Saidi et al.~\citep{Saidi_2014} was used to describe
the chemical bonding in the binary Al-Si system. An Al(Si) alloy layer
was deposited onto a Si(111) surface with either the unreconstructed
1$\times$1 or reconstructed 7$\times$7 atomic structure. The surface
area of the substrate was of 22$\times$19 nm$^{2}$ for the 7$\times$7
reconstructed surface and 42$\times$42 nm$^{2}$ for the 1$\times$1
reconstructed surface.

\begin{figure}
\centering \includegraphics[width=0.75\linewidth]{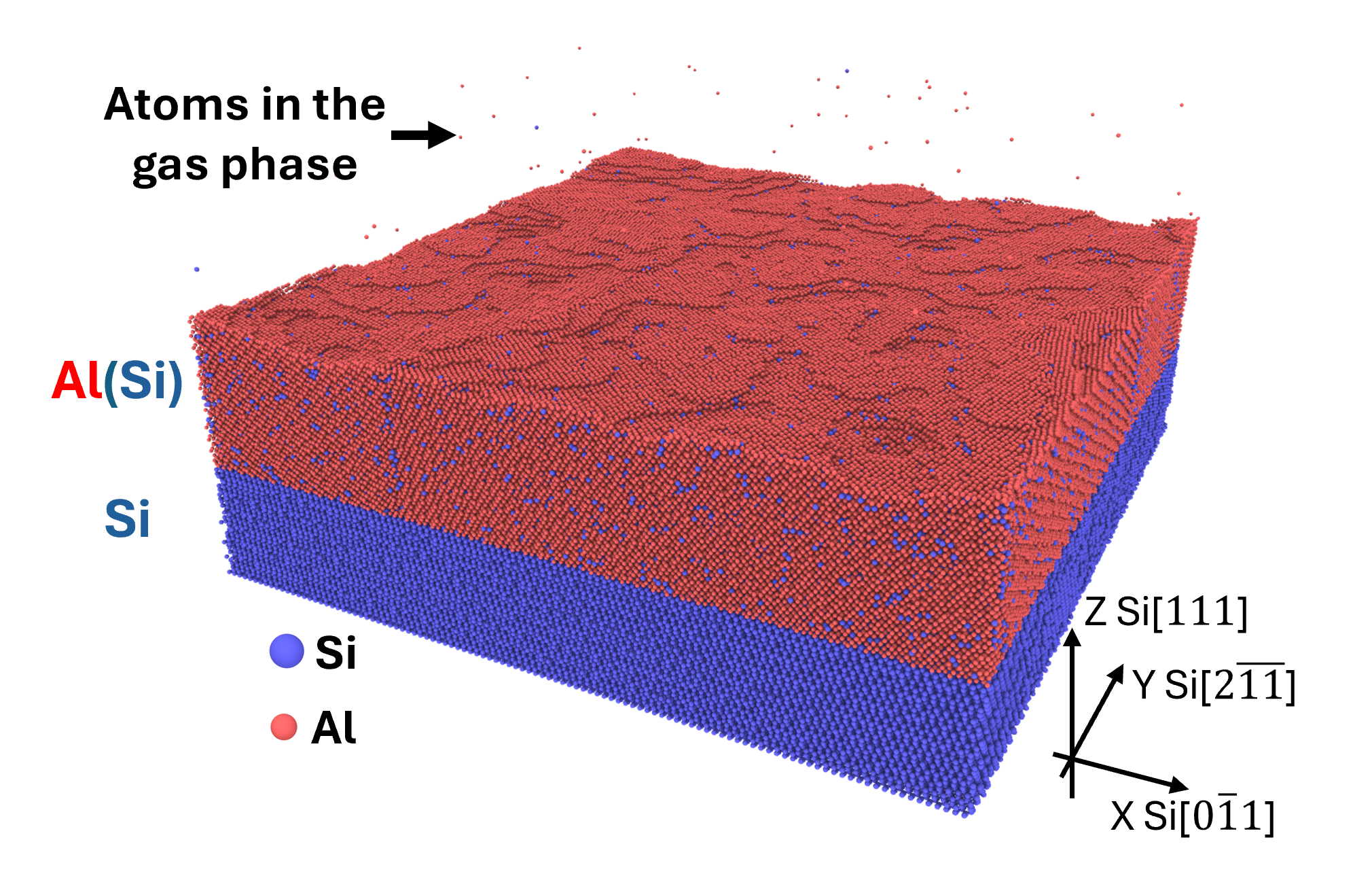}
\caption{Simulated growth of Al$_{0.93}$Si$_{0.07}$ layer on Si(111) substrate
with 7$\times$7 reconstructed surface by vapor deposition at the
temperature of 622 K. The Cartesian $X$, $Y$, and $Z$ directions
are along the $[0\overline{1}1]$, $[2\overline{1}\overline{1}]$,
and $[111]$ crystallographic directions of the substrate.}
\label{fig:Growth model}
\end{figure}

Figure \ref{fig:Growth model} explains the deposition setup using
an example of an Al$_{0.93}$Si$_{0.07}$ layer growing on the Si(111)
substrate at the growth temperature of 622 K. Prior to deposition,
the substrate had a 7$\times$7 reconstructed surface structure and
a thickness of 5 nm. It had a thermally expanded lattice parameter
corresponding to the growth temperature. The bottom two crystal planes
of the substrate were fixed. Periodic boundary conditions (PBCs) were
applied in the lateral directions $X$ and $Y$ with a fixed boundary
condition along the growth direction $Z$. Al and Si atoms with a
chosen Al:Si ratio were randomly created in a region (``gas'') well
above the substrate surface. The atomic velocities were towards to
the substrate and had magnitudes drawn from the Maxwell velocity distribution
at the growth temperature. The atoms hitting the substrate were gradually
building an Al(Si) alloy layer of the gas composition. This process
produced an Al(Si)/Si IPB with the chosen crystallographic orientation
of the substrate. Note that the simulation setup did not imposed any
particular crystallographic orientation of the Al(Si) layer. The growing
layer was free to choose the most favorite orientation during the
growth process. The alloy composition and temperature were chosen
so as to keep the alloy within the single-phase region on the left
of the solvus line on the Al-Si phase diagram \citep{Al-Si-IPB} computed
with the interatomic potential. 

We investigated the Al(Si)/Si IPBs at four temperatures: 578 K, 604
K, 622 K, and 638 K. These temperatures are close to the eutectic
temperature (approximately 677 K) predicted by the interatomic potential.
The corresponding solubility limits of Si in Al at these temperatures
are 3.5\%, 5.0\%, 6.8\%, and 8.0\%. (In this work, all chemical compositions
are measured in atomic per cents.) The deposition simulations targeted
chemical compositions close to the solubility limits at the respective
temperatures. The growth rate was around 0.1 monolayer per nanosecond
(ML/ns). Due to the time scale limitation of MD simulations, this
rate is several orders of magnitude higher than typical experimental
growth rates, which are several ML/s. To partially mitigate the effect
of the high growth rate, the first 2 MLs of the Al(Si) alloy were
deposited on the 1$\times$1 substrate with a relatively low rate
of 0.7 ML/$\mu\text{s}$ at temperatures ranging from 578 to 638 K.
For the 7$\times$7 reconstructed surface, the first 2 MLs were deposited
with the same 0.7 ML/$\mu\text{s}$ rate but at a higher temperature
of 770 K to accelerate the structural rearrangement to the 1$\times$1
reconstruction. The subsequent layers were grown at lower temperatures
from 578 to 638 K. Using a raised temperature is a common strategy
to mitigate the effects of a high growth rate \citep{gruber2017molecular,zhou2013stillinger,li2023dislocation}.
Applying a slow deposition for the initial layers is also beneficial
as the interface structure is largely determined during early stages
of the growth process.

\subsection{Diffusion coefficient calculations\label{subsec:D-calculation}}

In addition to Al(Si)/Si IPBs, a variety of microstructures formed
in the Al(Si) layer during the depositions. The diffusion coefficients
were computed for both the IPBs and these microstructures. Details
of the diffusion calculations are presented below.

The self-diffusion coefficients of Al and Si in Al(Si)/Si IPBs were
measured by tracking the motion of atoms within a 1.2 nm thick layer
centered at the interface during the MD simulations in the canonical
(NVT) ensemble. The 1.2 nm thickness approximately corresponds to
three atomic layers of Si and three atomic layers of Al(Si). The choice
of this thickness was based on the following. We computed the potential
energy of atomic layers as a function of distance normal to the interface
region and found that atoms within approximately 1.2 nm region centered
at the IPB had potential energy significantly different from that
in the bulk phases. The IPB position was determined by averaging the
$Z$-coordinates of the interface atoms having non-lattice environments.
The atomic environments were analyzed by the Polyhedral Template Matching
(PTM) algorithm in the Open Visualization Tool OVITO \citep{larsen2016robust}. 

The diffusion coefficients were calculated using the Einstein relations
$D_{x}=\langle x^{2}\rangle/2t$ and $D_{y}=\langle y^{2}\rangle/2t$,
where $\langle x^{2}\rangle$ and $\langle y^{2}\rangle$ are mean
squared displacements (MSDs) of atoms in the two in-plane directions
$X$ and $Y$, and $t$ is the simulation time. Only atoms that remained
in the interface region at the initial and final times were included
in the MSD calculation. In previous work \citep{Al-Si-IPB,Chesser:2022aa,Chesser:2024aa},
the diffusion coefficients were rescaled based on the fraction of
locally disordered atoms within the interface region. This rescaling
corrected for the possible underestimation of the diffusion coefficient
when the fixed-width interface layer contained not only high-diffusivity
pathways but also lattice-like atomic groups. Such groups did not
contribute much to the diffusion flux and were excluded from the diffusion
calculation. However, the IPBs studied in this work had very compact
and largely uniform atomic structures (see examples in Fig. \ref{fig:Interface structure})
that did not require such a rescaling. 

For interfaces obtained by Al(Si) deposition on the Si(111)-7$\times$7
surface, the Al and Si diffusion coefficients were below $10^{-13}\,\text{m}^{2}/\text{s}$,
which is close to the lower bound of measurable diffusivities in conventional
MD simulations. Therefore, we applied the parallel replica dynamics
(PRD) method \citep{Voter98,Perez:2015aa}, which is one of the accelerated
MD approaches \citep{Perez:2009aa}. The statistics of diffusive jumps
detected by PRD allowed us to quantify the diffusion coefficients.
Furthermore, the method identified typical atomic rearrangements causing
the atomic diffusion. The activation barriers of such rearrangements
were then computed by the nudged elastic band (NEB) method \citep{Henkelman00a,HenkelmanJ00}.
The PRD calculations provided insights into diffusion mechanisms in
the defect core regions, which would be difficult to obtain by other
methods. Technical details of the PRD calculations are summarized
in the Supplementary Information file accompanying this article.

Similar to the IPBs, self-diffusion coefficients of Al and Si in GBs
were computed by tracking the motion of atoms within a 2.0 nm probe
layer centered at the boundary during the MD simulations. The layer
was thicker than for the IPBs because the GBs had less ordered structures
with a thicker core region. After using the Einstein relations to
calculate the diffusion coefficients, the results were rescaled by
a factor equal to the ratio of the total number of atoms in the probe
layer to the number of locally disordered atoms identified by PTM
analysis in OVITO. The same methodology was used to compute the diffusion
coefficients of GB triple junctions (GBTJs) and IPB-GB triple lines
(GBTLs). For GBTJs, we tracked atoms in the common region of the three
GBs connected to the GBTJ. The region had a cylindrical shape with
a radius of 2 nm. For GBTLs, we tracked atoms in the common region
of the GB and the IPB, which had a rectangular shape with a height
of 1.2 nm ($Z$-direction) and a width of 2 nm (parallel to the interface).
The diffusion coefficients were calculated from the Einstein relation
and rescaled by the ratio of the total number of atoms in the region
to the number of locally disordered atoms.

In addition to IPBs, GBs, and their triple lines, threading dislocations
(TDs) formed in the growing Al(Si) layer during the deposition. Such
dislocations were connected to the IPB at one end and to the open
surface at the other end. To compute the diffusion coefficients in
the dislocation cores, we tracked the motion of atoms within a cylindrical
region whose axis was parallel to the $Z$ direction. The radius of
the cylinder was chosen to completely contain the TD core. As above,
the diffusion coefficients were extracted from the Einstein relation
followed by rescaling by the fraction of disordered atoms. 

For all diffusion coefficients calculated in this study, bootstrap
resampling of MSD curves was applied to compute error estimates. The
hyper-parameters chosen for this calculation were a block length of
20 ps and a number of resampled trajectories equal to 100.

\section{Results}

\subsection{Effect of Si(111) surface structure on the Al(Si)/Si(111) interphase
boundary}

The Si(111) surface structure can undergo several reconstructions
\citep{becker1985tunneling,binnig19837,brommer1992ab,smeu2012electronic,solares2005density,takayanagi1985structure}.
In this work, we investigated the two most common surface structures
corresponding to the 1$\times$1 and 7$\times$7 reconstructions.
For brevity, we refer to the IPBs obtained by Al(Si) deposition onto
the Si(111)-1$\times$1 and Si(111)-7$\times$7 substrates as (1$\times$1)
IPB and (7$\times$7) IPB, respectively. 

The simulations revealed two common features of the (1$\times$1)
and (7$\times$7) IPB cases. Firstly, the Al(Si) growth followed the
layer-by-layer mechanism in both cases. Secondly, the {[}111{]} crystallographic
direction in the deposited Al(Si) layer was always parallel to the
growth direction $Z$, regardless of the growth temperature or Si
concentration (Fig.~\ref{fig:Interface structure}). Thus, the obtained
orientation relationship across the IPB was ${\text{Al(Si)}\{111\}}\parallel{\text{Si}}\{111\}$.
However, differences were observed between the two IPB structures,
as discussed below.

\subsubsection{(1$\times$1) interphase boundary}

\begin{figure}
\centering \includegraphics[width=1\linewidth]{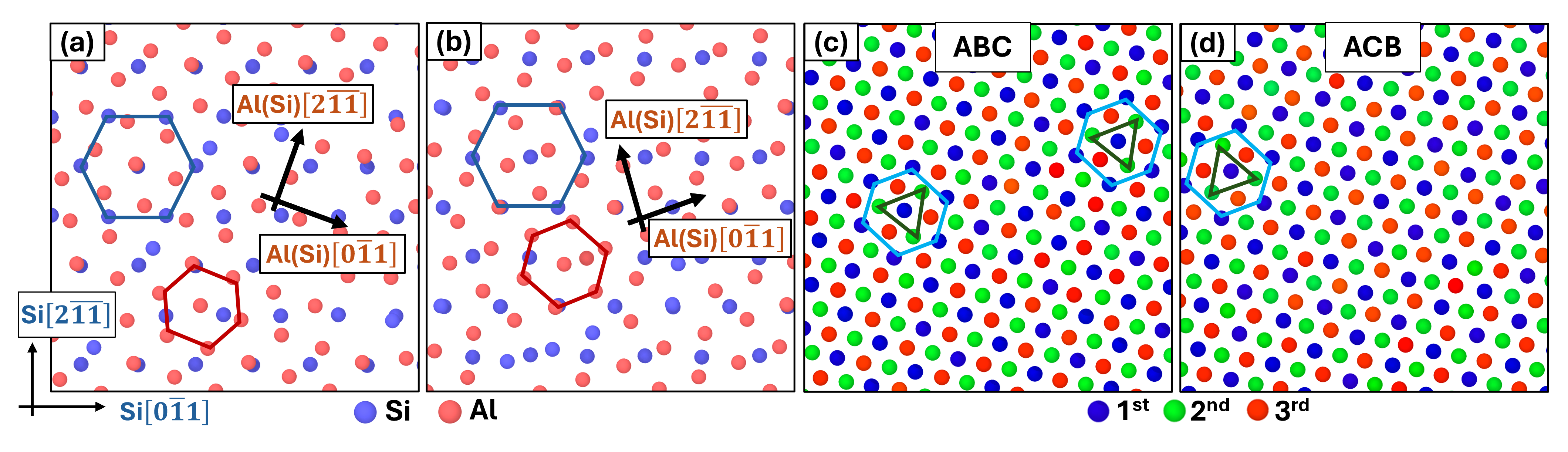}
\caption{Panels (a) and (b) are plane views of the (1$\times$1) IPB with two
different orientation relationships. The top atomic layer of the Si
substrate is superimposed on the bottom atomic layer of Al(Si). Panels
(c) and (d) show three sequential atomic layers of Al(Si) found in
different grains of the Al(Si) phase, illustrating two possible stacking
orders: A-B-C and A-C-B. The bottom plane A is in immediate contact
with the Si substrate. The atoms are colored according to their plane.
The hexagons outline the structural motifs.}
\label{fig:Alignments and stacking order}
\end{figure}

The (1$\times$1) IPBs were found to be atomically sharp with very
little intermixing between Al and Si. Simulations revealed two possible
in-plane orientations at (1$\times$1) IPBs. Namely, the $[0\overline{1}1]$
direction in the Al(Si) layer was rotated either +$19^{\circ}$ or
$-19^{\circ}$ away from the $[0\overline{1}1]$ direction of the
Si substrate. Different grains in Al(Si) exhibited one of these two
orientations, as illustrated in Fig.~\ref{fig:Alignments and stacking order}.
Each of the two orientations could additionally exhibit two different
stacking sequences of the Al(Si) planes at the interface: either A-B-C
or A-C-B. While both sequences represent the FCC structure, they were
rotated relative to each other by about $70.5^{\circ}$ about the
common tilt axis {[}110{]}. The two in-plane rotations and the two
stacking orders gave rise to four different grain orientations of
the Al(Si) layer, which were all observed in the deposition simulations. 

\begin{figure}
\centering \includegraphics[width=1\linewidth]{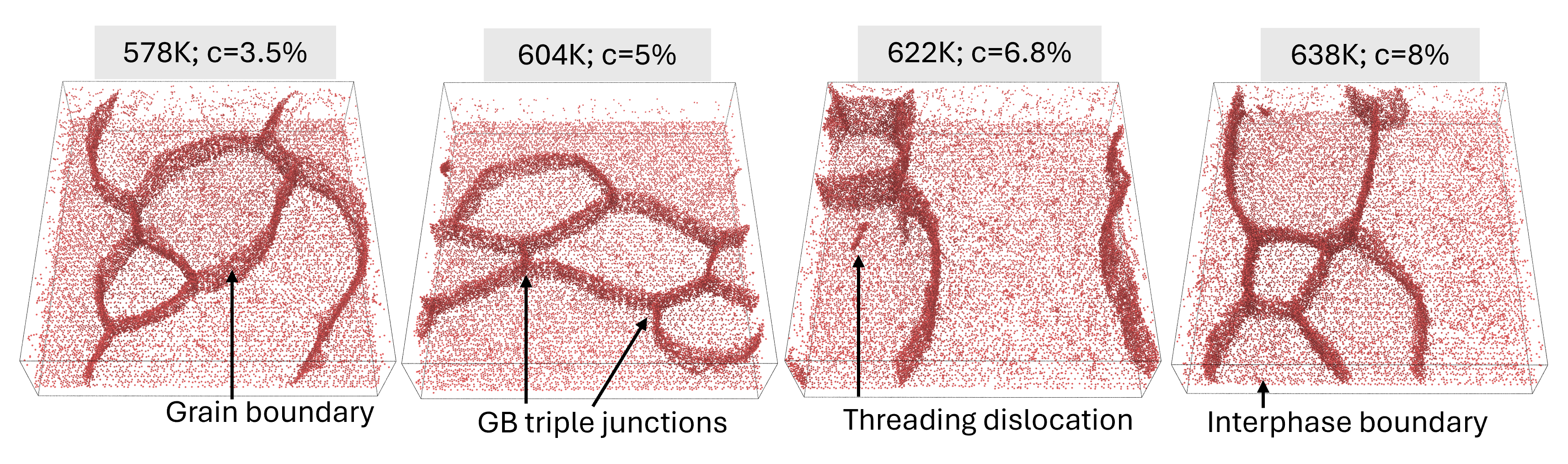}
\caption{Perspective view of the Al(Si) layer deposited on Si(111)-1$\times$1
substrate at four temperatures ranging from 578 K to 638 K and Si
concentrations $c$ between 3.5\% to 8.0\%. Only atoms in locally
disordered environments are shown to reveal IPBs, GBs, GBTLs, GBTJs,
and threading dislocations.}
\label{fig:AlSi structures}
\end{figure}

\begin{figure}
\centering \includegraphics[width=0.9\linewidth]{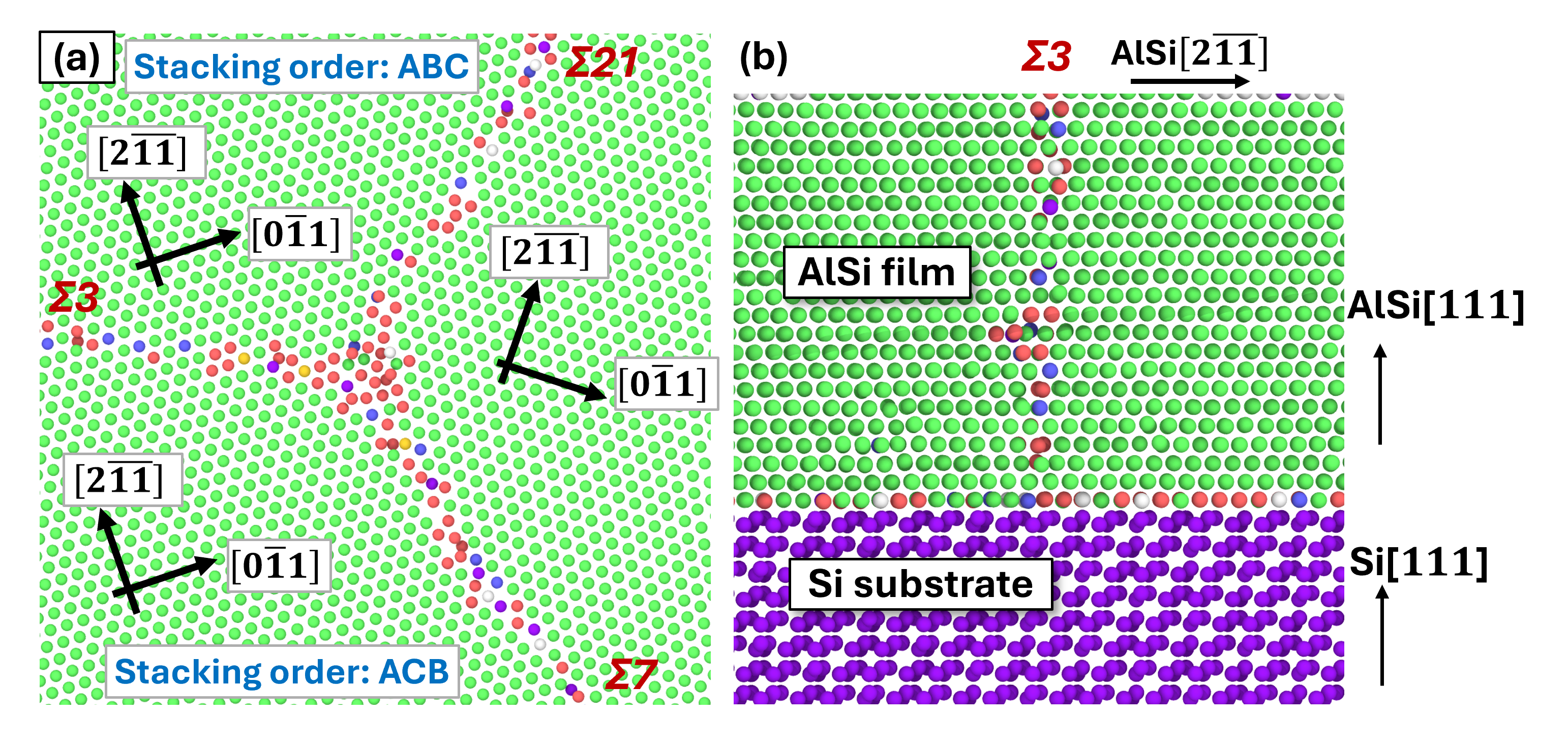}
\caption{(a) Plane view of three grains in Al(Si) layer with the Si concentration
of 8\% deposited on Si(111)-1$\times$1 substrate at 638 K. The GBs
and their triple junction are normal to the page. (b) Side view of
this structure showing the $\Sigma$3 incoherent twin boundary. The
atoms are colored by local coordination number according to the scheme:
green for FCC, purple for simple cubic, and all other colors represent
atoms in locally disordered environments. The atoms at GBs and the
IPB are red, grey, or blue.}
\label{fig:GB3}
\end{figure}

Fig.~\ref{fig:AlSi structures} shows typical polycrystalline structures
of the Al(Si) layers deposited on Si(111)-1$\times$1. The structures
are composed of columnar grains with the common {[}111{]} axis normal
to the substrate. While specific grain shapes and sizes depended on
the deposition temperature and chemical composition, all structures
contained GBs, triple-line defects, and dislocations. Three types
of GBs were observed: $\Sigma$3 $\langle110\rangle$ $70.5^{\circ}$
tilt GB (incoherent twin boundary), $\Sigma$7 $\langle111\rangle$
$38.2^{\circ}$ tilt GB, and $\Sigma$21 $\langle111\rangle$ $21.8^{\circ}$
tilt GB. As an example, in Fig.~\ref{fig:GB3}(a) we show a GBTJ
connected to $\Sigma$3, $\Sigma$7, and $\Sigma$21 GBs. The GBTJ
is normal to the page, while the three GBTLs (GB/IPB intersections)
are revealed as loci of non-FCC atoms. The $\Sigma$3 incoherent twin
GB is formed between two grains with similar in-plane orientations
but different stacking orders. A side view of this boundary is shown
in Fig.~\ref{fig:GB3}(b).

We were unable to compare the simulation results for the (1$\times$1)
IPBs with experimental observations. To our knowledge, epitaxial growth
of Al on Si(111)-1$\times$1 is difficult to implement. High-temperature
annealing to remove the surface oxide renders the Si(111)-1$\times$1
surface unstable and causes its reconstruction to other surface structures
\citep{mcskimming2017metamorphic}. This explains the lack of clear-cut
experimental results for the (1$\times$1) IPB structures. 

\subsubsection{(7$\times$7) interphase boundary}

\begin{figure}
\centering \includegraphics[width=0.9\linewidth]{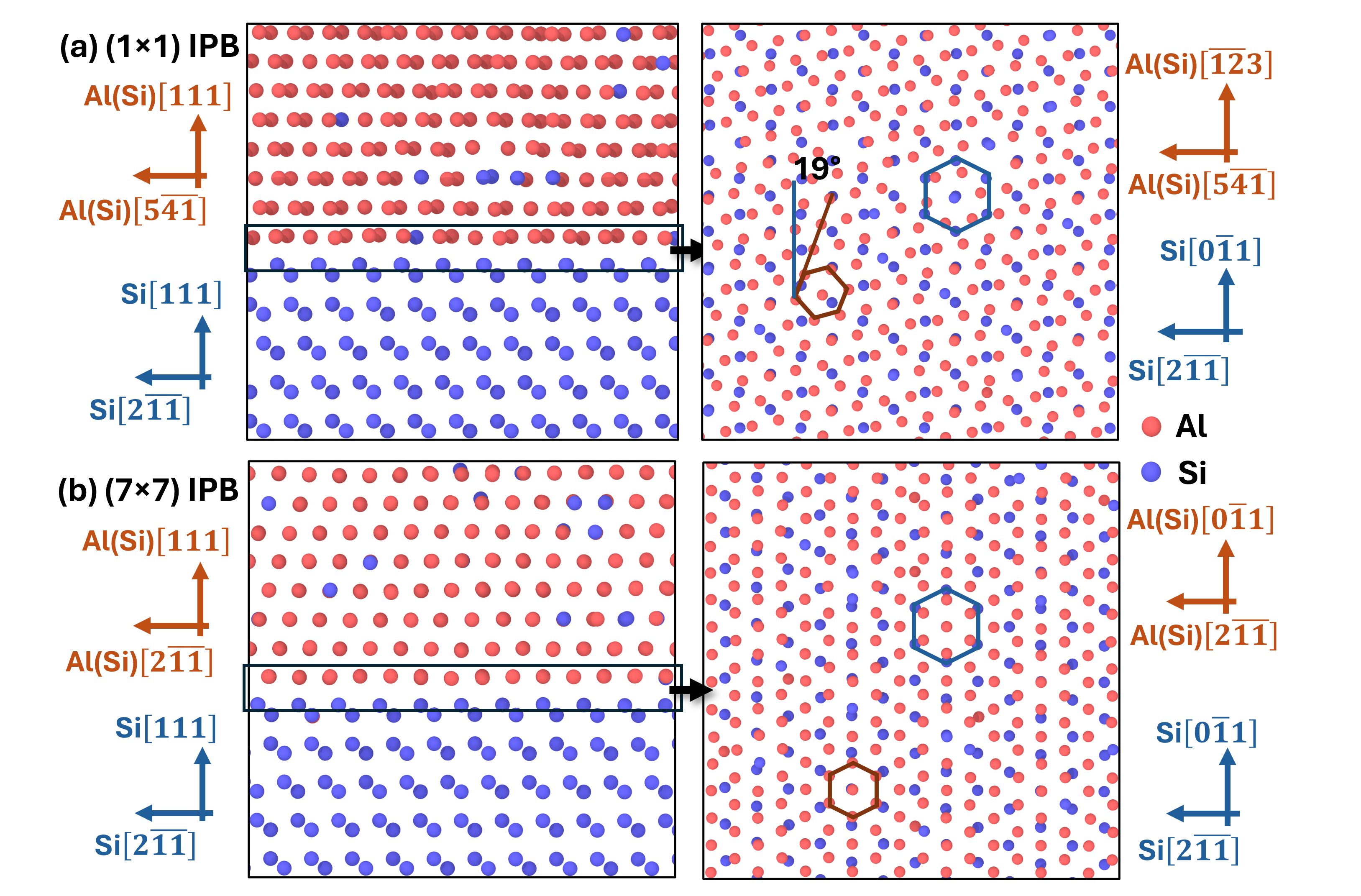}
\caption{Structures of Al$_{0.92}$Si$_{0.08}$(111)/Si(111) interphase boundaries
obtained by vapor deposition at 638 K. Two orientation relationships
and interface structures were obtained for different Si surface reconstructions:
(a) (1$\times$1) surface and (b) (7$\times$7) surface. The left
column shows side views of the interface structures. The right column
shows a plane view of the top atomic layer of the Si substrate and
the bottom atomic layer of Al(Si). Atoms in red and blue represent
Al and Si atoms, respectively. The hexagons outline structural motifs.}
\label{fig:Interface structure}
\end{figure}

The (7$\times$7) IPB exhibits a cube-on-cube orientation relationship,
in which the Al(Si) layer has its $[0\overline{1}1]$ and $[2\overline{1}\overline{1}]$
directions parallel to the $[0\overline{1}1]$ and $[2\overline{1}\overline{1}]$
directions of the Si substrate (Fig.~\ref{fig:Interface structure}(b)).
The deposition temperature has little effect on the interface structure.
Simulations show that the Si(111)-7$\times$7 surface undergoes a
structural rearrangement during the Al(Si) deposition. Specifically,
the dimer-adatom-stacking fault structure of this surface \citep{takayanagi1985structure}
transforms into the Si(111)-1$\times$1 structure. This results in
the orientation relationship ${\text{Al(Si)}\{111\}}\parallel{\text{Si}}\{111\}$
with ${\text{Al(Si)}}\langle110\rangle\parallel{\text{Si}}\langle110\rangle$.
Note that this orientation relationship differs from the one obtained
by depositing Al(Si) directly on the Si(111)-1$\times$1 surface,
in which case the ${\text{Al(Si)}}\langle110\rangle$ and ${\text{Si}}\langle110\rangle$
axes are misaligned by $\pm19^{\circ}$ (see above). Another distinct
feature of the Si(111)-7$\times$7 deposition is some intermixing
between the Al and Si atoms in the interface region, which was not
observed in the (1$\times$1) IPBs. The intermixed Al atoms are constrained
to the first layer of the Si substrate adjacent to the IPB. This intermixing
is coupled to the structural rearrangement occurring on the surface
of the Si substrate. (The described intermixing is not apparent in
Fig.~\ref{fig:Interface structure}(b) because it only shows a small
interface area.) 

The simulation results agree well with the experimental interface
structures obtained by growing Al thin films on Si(111)-7$\times$7
substrates by molecular beam epitaxy \citep{mcskimming2017metamorphic,liu2018perfect}.
The epitaxy experiments reveal the same orientation relationship and
a similar structural transformation of the Si surface as found in
the simulations. Both simulations and experiments indicate that the
initial Si surface reconstruction significantly impacts the orientation
relationship across the Al(Si)/Si interface obtained by vapor deposition.

The simulations have also shown that the Al(Si) layers deposited on
the Si(111)-7$\times$7 substrates are single-crystalline. This observation
is consistent with the experimental observation of very low GB density
(significantly less than 1 $\mu\text{m}^{-1}$) in Al layers deposited
on Si(111)-7$\times$7 substrates \citep{liu2018perfect}. Such a
low GB density is unlikely to be captured by MD deposition simulations,
which are limited to the $\sim10$ nm length scale. 

\subsection{Diffusion coefficients from conventional MD simulations}

Next, we present the calculation results for Al and Si diffusion along
the IPBs and other defects in the microstructure of the Al(Si)/Si
system. These results were obtained by conventional MD simulations.

\begin{figure}
\centering \includegraphics[width=0.8\linewidth]{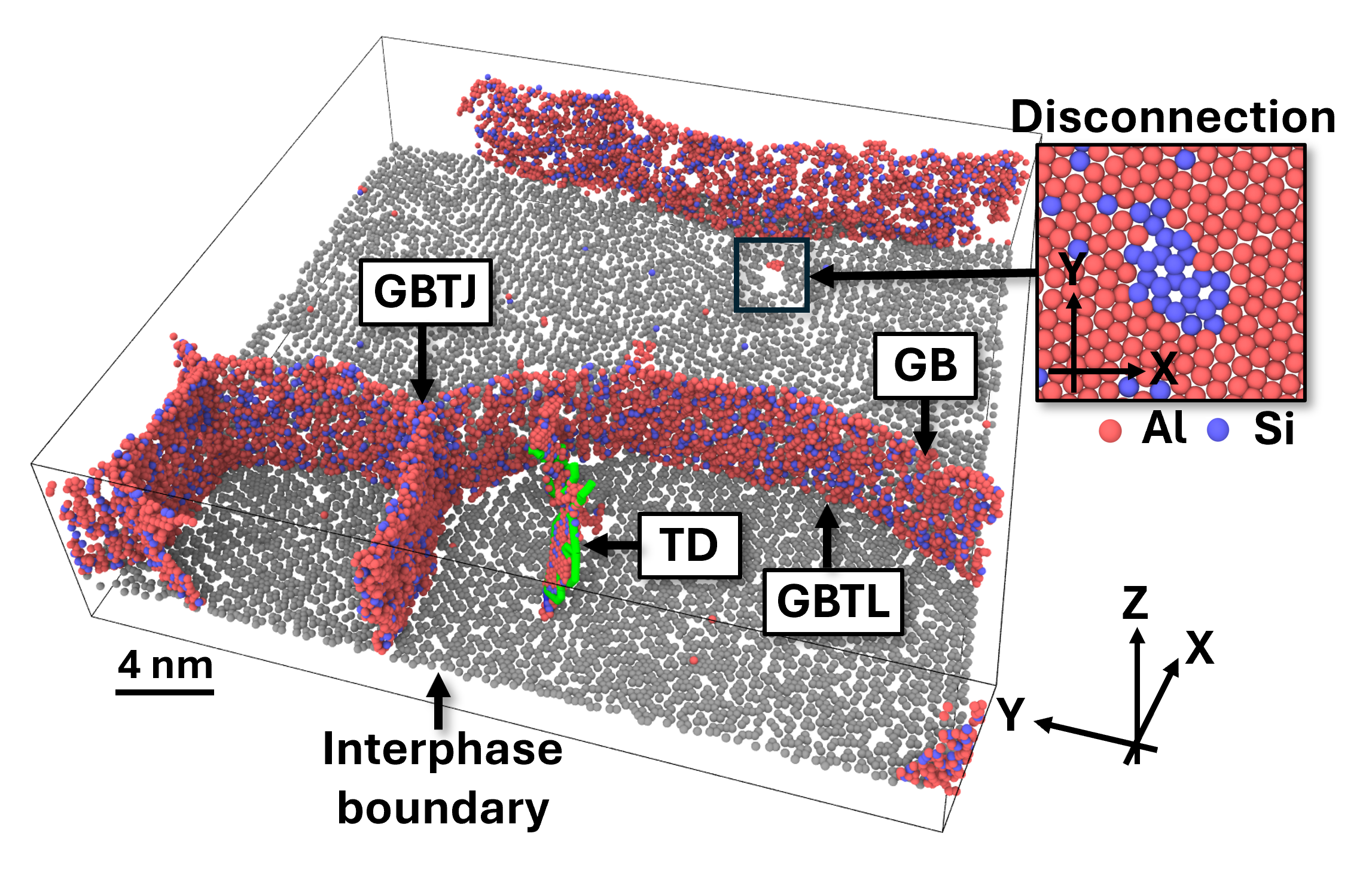}
\caption{Perspective view of a polycrystalline Al$_{0.93}$Si$_{0.07}$ layer
deposited on a Si(111)-1$\times$1 substrate. The arrows point to
defects used in diffusion calculations: interphase boundary (IPB),
grain boundaries (GBs), grain boundary triple junctions (GBTJ), GB-IPB
triple lines (GBTL), and threading dislocations (TD). The zoomed-in
image shows a disconnection (DC) loop at the IPB, with the blue and
red colors representing Si and Al atoms, respectively.}
\label{fig:micrstructures}
\end{figure}

\begin{figure}
\noindent \begin{centering}
\includegraphics[width=1\textwidth]{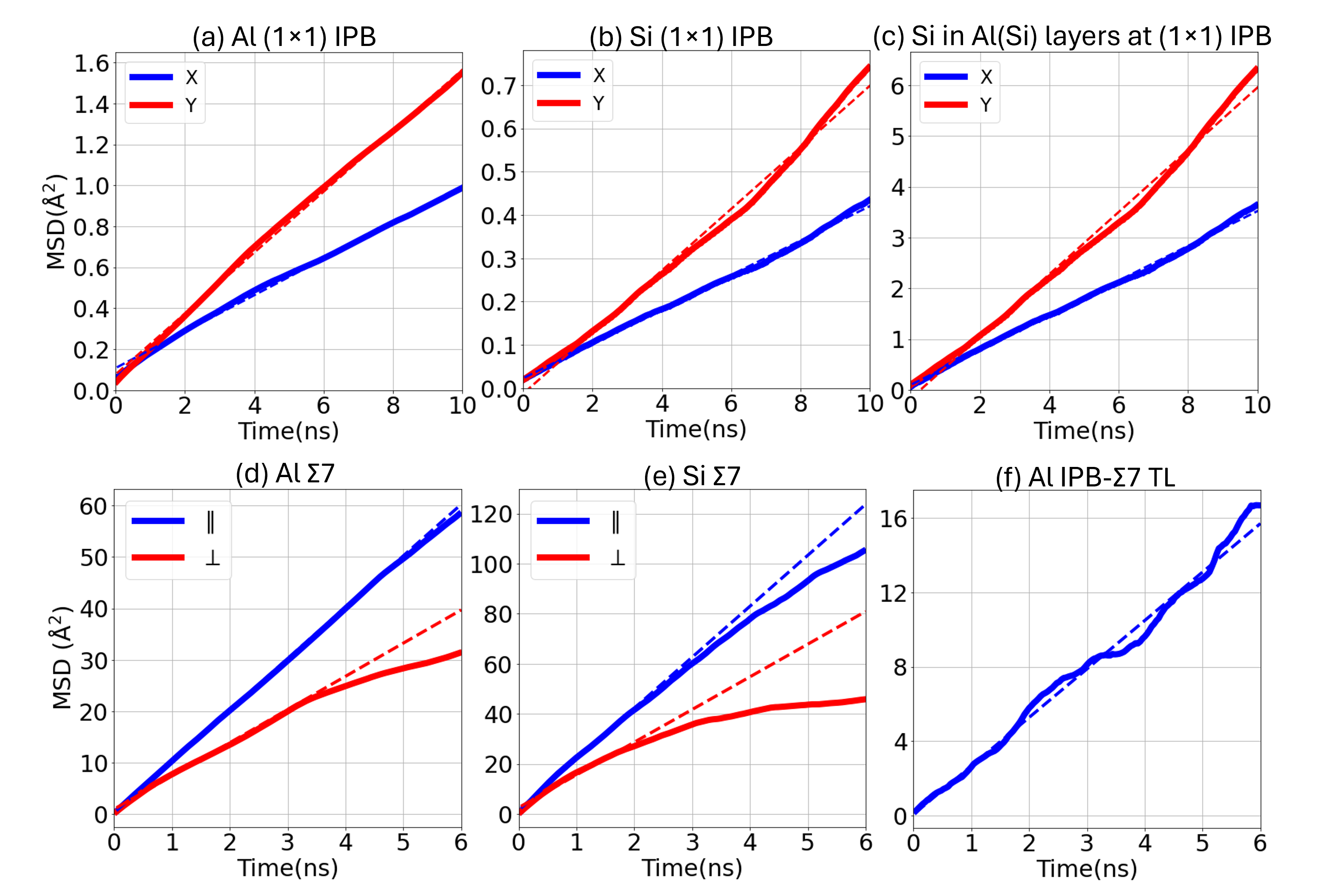}
\par\end{centering}
\caption{Representative MSD versus time plots obtained by MD simulations. (a,b)
Al and Si atoms diffusing in the $X$ and $Y$ directions in the $(1\times1)$
IPB. (c) Si atoms diffusing in the $(1\times1)$ IPB excluding Si
atoms in the substrate. (d,e) Al and Si atoms in a $\Sigma7$ GB segment
diffusing in directions parallel ($\Vert$) and normal ($\bot$) to
the substrate. (f) Al atoms diffusing along a triple line between
the $(1\times1)$ IPB the $\Sigma7$ GB. \label{fig:MSD-plots}}

\end{figure}

Fig.~\ref{fig:micrstructures} shows a typical microstructure formed
by deposition on the Si(111)-1$\times$1 substrate. The diffusion
coefficients were computed using the methodology discussed in Section
\ref{subsec:D-calculation}, which is based on Einstein's formula
for MSDs of atoms. Typical MSD versus time plots are shown in Fig.~\ref{fig:MSD-plots}.
In most cases, the predicted linear relation is followed fairly well,
although deviations from linearity were also observed. At least two
factors could be responsible for such deviations. The first factor
is the spatial heterogeneity of the diffusion pathways. For example,
atoms near the center of the interface core usually possess a higher
mobility than atoms in distorted lattice planes immediately adjacent
to the interface. Since both are included in the MSD calculation,
deviations from the Einstein relation (which assumes a homogeneous
medium) are inevitable. They can accumulate with time, leading to
a nonlinear behavior at longer times. The second factor is that the
MSD calculation only includes atoms that remained within the probe
region during the chosen time interval. Higher-mobility atoms are
more likely to leave the probe region and not contribute to the MSD.
This leads to underestimated MSDs, causing downward deviations from
the linearity of the MSD-time plots in the long-time limit. When a
plot deviated from linear behavior, we used its initial (short-time)
part to extract the diffusion coefficient.

\begin{figure}
\centering \includegraphics[width=1\textwidth]{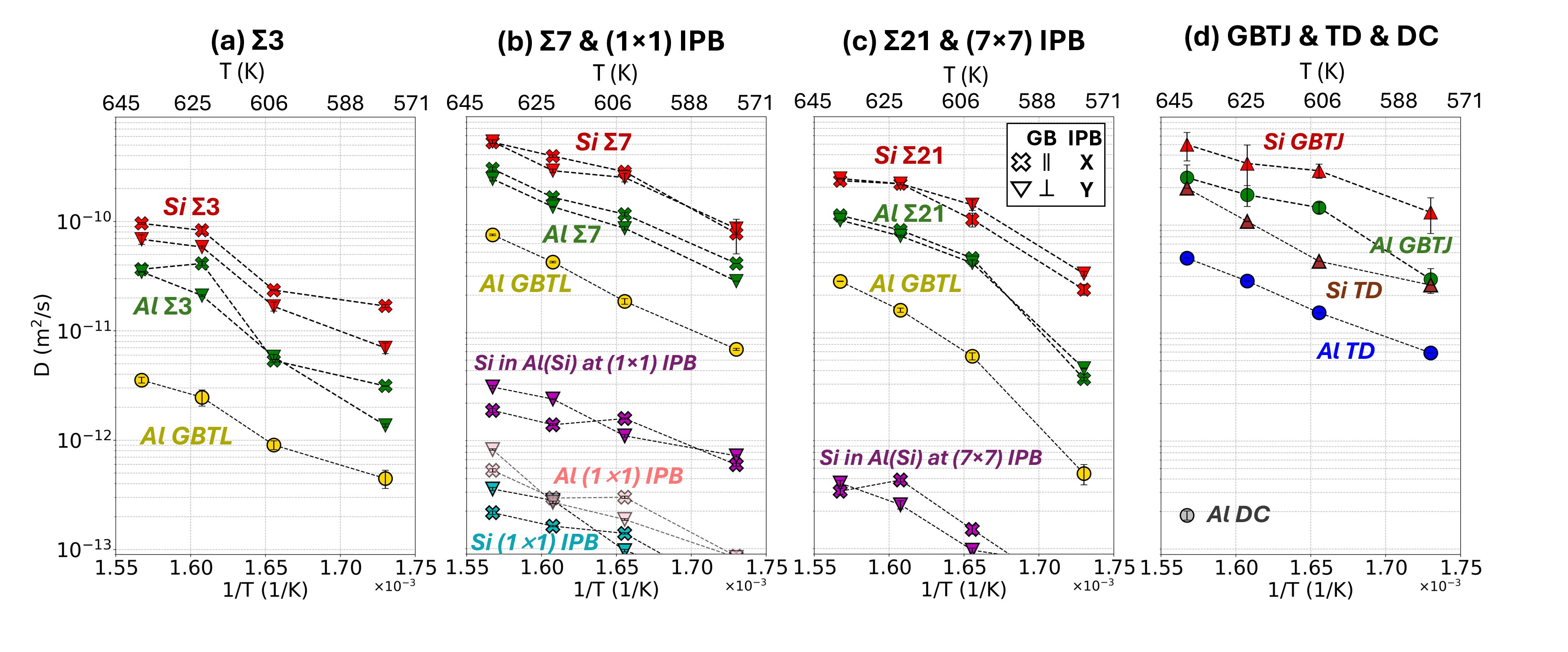}
\caption{Arrhenius diagrams of Al and Si diffusion at the (a) $\Sigma$3 GB,
(b) $\Sigma$7 GB and (1$\times$1) IPB, (c) $\Sigma$21 GB and (7$\times$7)
IPB. The Al diffusion coefficient at GBTLs is included for comparison.
(d) Al and Si diffusion at GBTJs, TDs, and DC loops in the (1$\times$1)
IPB. The $X$ and $Y$ directions correspond to the $[0\overline{1}1]$
and $[2\overline{1}\overline{1}]$ directions of the Si substrate.
The symbols $\parallel$ and $\perp$ refer to directions in the GB
plane parallel and perpendicular to the IPB, respectively. For the
$\Sigma$7 and $\Sigma$21 GBs, the $\perp$ direction is aligned
with the {[}111{]} tilt axis.}
\label{fig:Arrhenius diagrams}
\end{figure}

In Fig.~\ref{fig:Arrhenius diagrams}, we present the Arrhenius diagrams
(log diffusivity versus inverse temperature) for the Al and Si diffusion
coefficients in the defect structures. Recall that at each temperature,
the chemical composition of the Al(Si) alloy corresponds to the thermodynamic
equilibrium between the phases at that temperature. In other words,
the chemical composition varies with temperature. The main results
of the calculations are summarized below.

For IPBs and GBs, the diffusivity tends to decrease in the order:
(Si in $\Sigma$7 GB) $>$ (Si in $\Sigma$21 GB) $>$ (Al in $\Sigma$7
GB) $>$ (Al in $\Sigma$21 GB) $\approx$ (Si in $\Sigma$3 GB) $>$
(Al in $\Sigma$3 GB) $>$ (Si in Al(Si) at (1$\times$1) IPB) $>$
(Al in (1$\times$1) IPB) $>$ (Si in (1$\times$1) IPB) (Fig.~\ref{fig:Arrhenius diagrams}).
Overall, GBs exhibit much higher diffusivity than IPBs, with the difference
reaching about two orders of magnitude. Among the GBs, the $\Sigma$7
GB has the highest diffusivity and the $\Sigma$3 GB the lowest. This
trend correlates with the strength of Si GB segregation: the $\Sigma$7
GB exhibits the highest Si segregation and the $\Sigma$3 GB the lowest.
(The results for Si GB segregation are presented in the Supplementary
Information file.) The diffusion coefficients in the (1$\times$1)
IPBs are near the lower limit of diffusion coefficients measurable
by conventional MD simulations (around $10^{-13}\,\text{m}^{2}/\text{s}$).
The diffusion coefficients of Si and Al in the (7$\times$7) IPB are
below this limit and are not shown in Fig.~\ref{fig:Arrhenius diagrams}. 

Si exhibits a higher GB diffusivity than Al, which is consistent with
previous work \citep{Al-Si-IPB,Chesser:2024aa}. In contrast, Si diffusivity
in the IPBs is consistently lower than Al diffusivity. This trend
was also noted in \citep{Al-Si-IPB,Chesser:2024aa} and was explained
by the ordering effect imposed by the Si substrate. Indeed, at the
temperatures studied here, the Si atoms in the substrate are virtually
immobile. Their highly ordered spatial arrangement imposes a period
constraint on the IPB atoms, reducing their mobility. To verify this
effect, we recomputed the diffusion coefficients of Si in the (1$\times$1)
and (7$\times$7) IPBs by excluding the substrate Si atoms from the
probe layer. Only Si atoms diffusing on the Al(Si) side of the interface
region were included in the MSD calculations. The Si diffusion coefficients
obtained were still much lower than those for GB diffusion but an
orderer of magnitude higher than the diffusion coefficients computed
by including all atoms within the probe layer (Fig.~\ref{fig:Arrhenius diagrams}(b,c)).
Furthermore, as shown on the Arrhenius diagram, Si diffusivity on
the Al(Si) side of the IPBs is consistently higher than the Al diffusivity,
which aligns with the proposed explanation.

The anisotropy of GB and IPB diffusion is relatively small. In most
cases, it cannot be detected beyond the statistical scatter of the
data points. The only cases where the anisotropy is likely to be statistically
significant is for Si diffusion in the $\Sigma3$ and $\Sigma$21
GBs.

Si diffuses in GBTJs faster than Al, and both diffusivities are close
to that in $\Sigma$7 GBs (Fig. \ref{fig:Arrhenius diagrams}d). This
observation suggests that the GBTJ diffusivity is governed by the
highest-diffusivity GB connected to the junction. Al diffusivity in
GBTLs correlates with the respective GB diffusivity and decreases
in the order ($\Sigma$7 GB) $>$ ($\Sigma$21 GB) $>$ ($\Sigma$3
GB). Si diffusion in GBTLs is too slow to be quantified because many
Si atoms in the GBTL belong to the top substrate layer and are virtually
immobile. Al diffusion along disconnection loops in the (1$\times$1)
IPB is on the order of $10^{-13}\,\text{m}^{2}/\text{s}$ (Fig.~\ref{fig:Arrhenius diagrams}(d)),
while Si diffusion is too slow to be measured. Thus, diffusion in
the (1$\times$1) IPB is dominated by GBTLs. 

\subsection{Simulations by parallel replica dynamics}

\subsubsection{Investigation of diffusion mechanisms}

As discussed above, calculations of the diffusion coefficients in
the (7$\times$7) IPB are below the capabilities of conventional MD
simulations. The same is true for diffusion in the (1$\times$1) IPB
outside the GBTLs and disconnection loops. To overcome this limitation,
we applied the PRD method \citep{Voter98,Perez:2015aa} at temperatures
ranging from 578 K to 638 K to estimate the IPB diffusivities and
gain insights into the diffusion mechanisms. 

\begin{figure}
\noindent \centering{}\centering \includegraphics[width=0.8\linewidth]{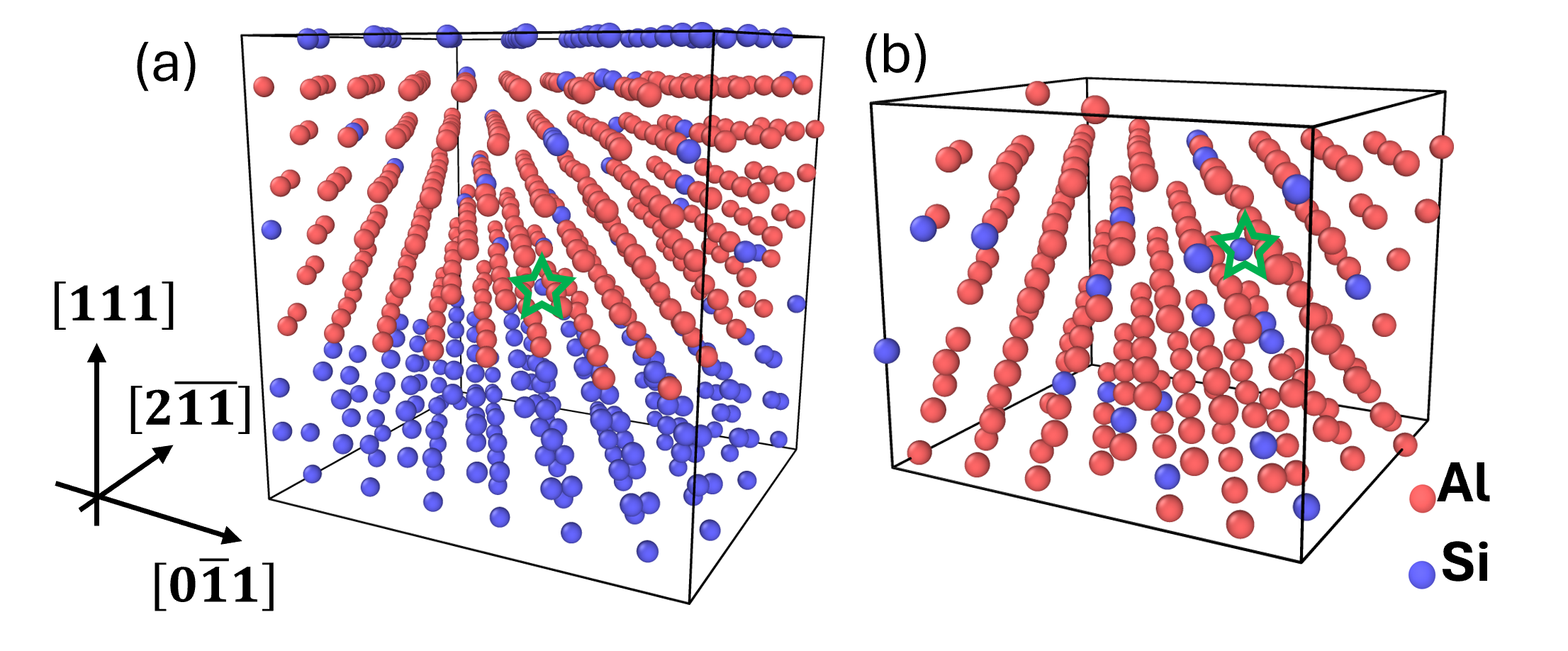}
\caption{Small models prepared for PRD simulations. (a) Al(Si)/Si model representing
a (7$\times$7) IPB. (b) Single-crystalline Al(Si) model. The pentagrams
mark Si interstitials.}
\label{fig:PRD models}
\end{figure}

The deposited Al(Si)/Si systems were too large for PRD simulations.
Instead, we constructed a set of relatively small fully periodic models
mimicking the deposited structures. To model the (7$\times$7) IPB,
the $[0\overline{1}1]$ and $[2\overline{1}\overline{1}]$ directions
in the Al(Si) layer were aligned parallel to those in the substrate
(Fig.~\ref{fig:PRD models}(a)). Due to the PBCs, the model effectively
contained two IPBs. The substrate contained 216 Si atoms and the Al(Si)
layer contained 384 atoms. To represent the (1$\times$1) IPB, the
$[0\overline{1}1]$ and $[2\overline{1}\overline{1}]$ directions
in Al(Si) we misaligned by approximately $19^{\circ}$ relative to
those in the Si substrate. In this case, the substrate contained 288
Si atoms and the Al(Si) layer contained 504 atoms. In both cases,
the system dimensions and the numbers of atoms were adjusted to satisfy
the PBCs. For comparison, we also constructed a set of single-phase
Al(Si) structures containing 216 atoms, as illustrated in Fig.~\ref{fig:PRD models}(b).
In all cases, the Si concentration in Al(Si) ranged from approximately
3.5\% to 8\%, matching the estimated solid solubility of Si in Al
along the solvus line on the phase diagram. 

The initial structures did not contain any point defects, and PRD
detected very few diffusive events. This is not surprising because
the deposited structures contained vacancies and interstitial atoms
responsible for diffusion activity. Therefore, we next introduced
an interstitial or a vacancy in the models. The results are summarized
below.

\begin{figure}
\centering \includegraphics[width=0.8\linewidth]{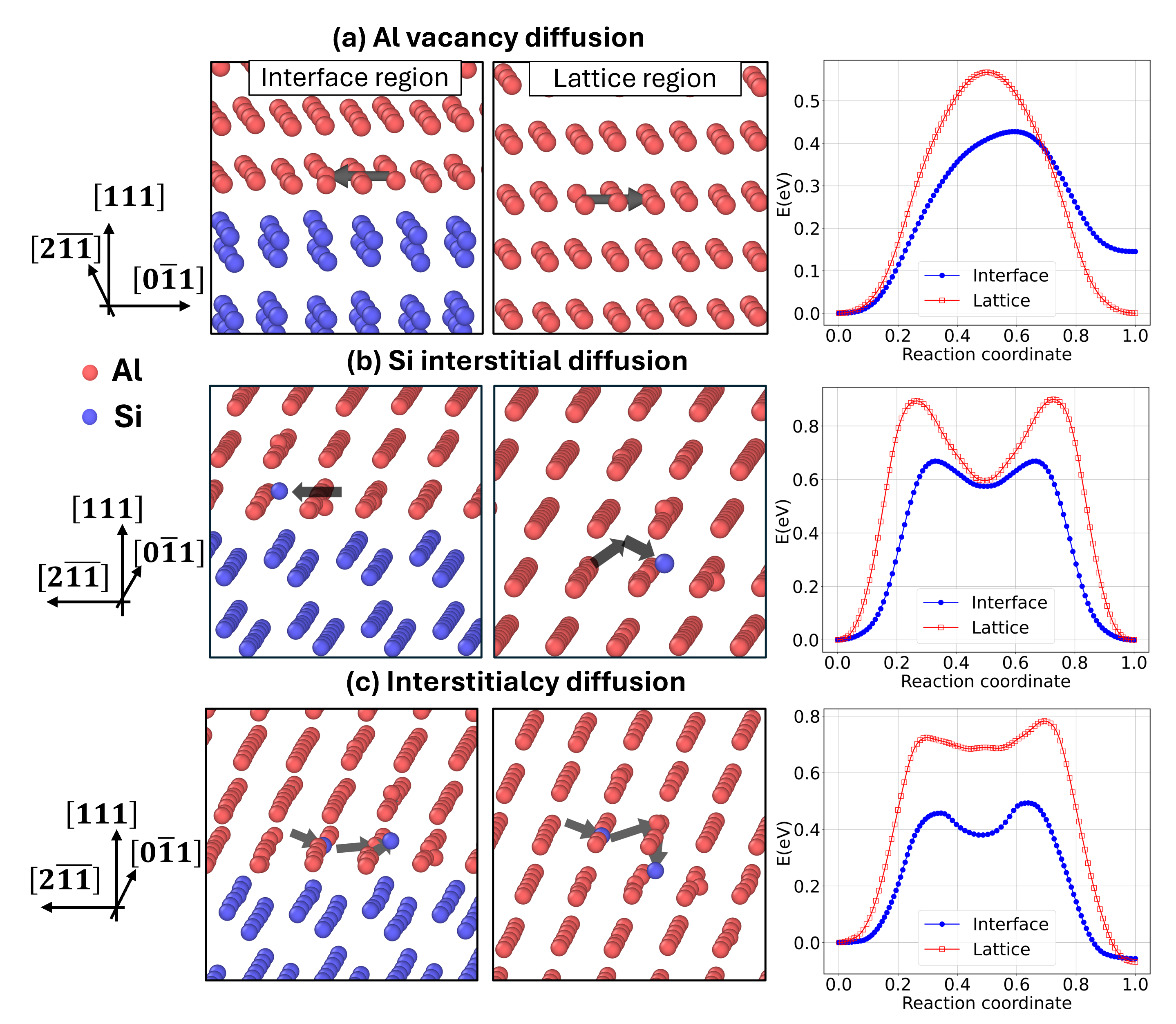}
\caption{Illustration of diffusion mechanisms reveled by PRD simulations. (a)
Al vacancy diffusion. (b) Direct Si interstitial mechanism. In the
lattice, the Si interstitial jumps between tetrahedral sites. (c)
Interstitialcy mechanism involving Si and Al atoms in the (7$\times$7)
interface and a lattice region. The arrows present atomic displacements.
The third column presents selected minimum-energy paths obtained by
NEB calculations.}
\label{fig:NEB results}
\end{figure}

An Al vacancy diffuses primarily by exchanges with Al atoms, regardless
of the Si concentration in Al(Si) (Fig. \ref{fig:NEB results}a).
Even in systems containing 8\% of Si in Al(Si), PRD detected over
a thousand vacancy-Al exchanges and only a few vacancy exchanges with
Si atoms. Very few Al self-interstitials were observed. When an interstitial
Al atom was introduced into the model prior to the PRD simulation,
it typically kicked a substitutional Si atom out of its position and
took its place, creating a Si interstitial.

The diffusion behavior of Si atoms is more complex. Si interstitial
atoms in Al(Si) were found in three different configurations: $\langle100\rangle$
dumbbell (two Si atoms aligning parallel to the $\langle100\rangle$
crystallographic direction sharing a lattice site), tetrahedral configuration,
and octahedral configuration. Octahedral interstitials were observed
least frequently. Si atoms diffused by either the direct interstitial
mechanism or the indirect mechanism called the interstitialcy mechanism.
In the first case, the interstitial Si atom hops directly to a neighboring
interstitial position without displacing lattice atoms. Most of the
time, the atom hops between tetrahedral interstitial sites (Fig. \ref{fig:NEB results}b).
Note that only Si atoms participate in this process, causing no Al
diffusion. In the interstitialcy events, multiple atoms move simultaneously,
forming a structural excitation causing a collective string-like atomic
displacement of several atoms (Fig.~\ref{fig:NEB results}c). The
event typically, starts with a Si interstitial atom jumping into a
neighboring lattice site, displacing the resident atom into an adjacent
interstitial position. The displaced atom, in turn, displaces another
atom from its substitutional position, and the displacement process
continues. This chain of displacements typically ends with the formation
of a new Si interstitial. Note that both Al and Si atoms can participate
in the interstitialcy mechanism, causing diffusion of both species.
With increasing Si concentration in Al(Si), the Si interstitials are
more likely to diffuse by the interstitialcy mechanism rather than
the direct interstitial mechanism in both the interface regions and
the lattice. 

Fig.~\ref{fig:NEB results} illustrates the diffusion mechanisms
mentioned above and the respective NEB minimum energy paths in the
(1$\times$1) and (7$\times$7) IPB regions and the Al(Si) lattice.
In all three cases, the diffusion mechanisms are qualitatively similar.
Thus, we do not find a unique mechanism of IPB diffusion. However,
the energy barriers in the interface region are lower than in the
lattice. The energy difference ranges from a few hundredths to a few
tenths of eV and is responsible for the accelerated diffusion in the
interfaces relative to the lattice. 

\subsubsection{Diffusion coefficients from parallel replica dynamics}

\begin{figure}
\centering \includegraphics[width=0.75\linewidth]{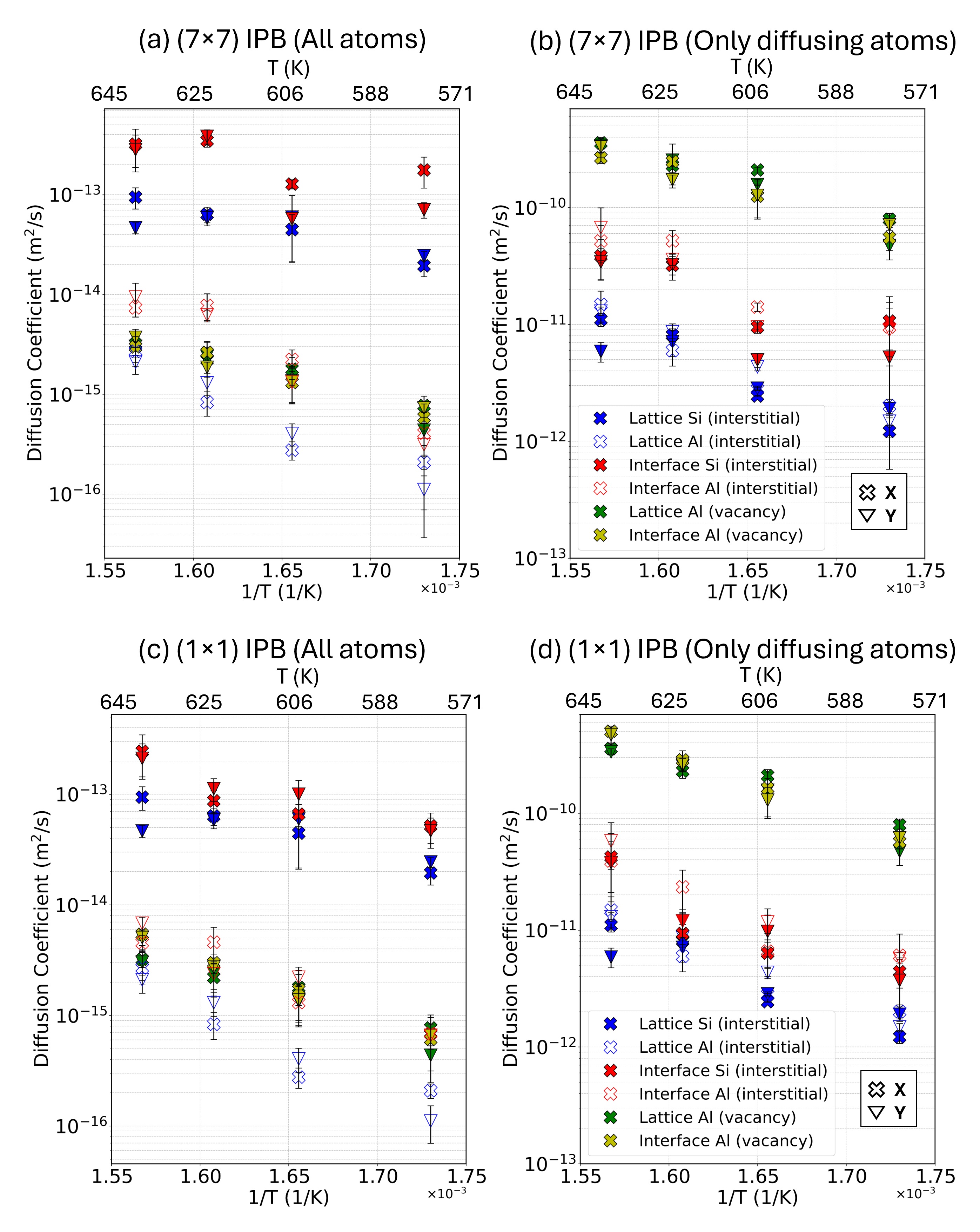}
\caption{Diffusion coefficients for Al and Si atoms obtained by PRD calculations
considering all atoms (a,c) and only diffusing atoms (b,d). The results
for the (7$\times$7) IPB and (1$\times$1) are shown in (a,b) and
(c,d), respectively.}
\label{fig:PRD diffusion}
\end{figure}

Figure \ref{fig:PRD diffusion} shows the Al and Si diffusion coefficients
obtained by the PRD simulations. Since the diffusion coefficients
in the (7$\times$7) and (1$\times$1) IPBs (unaffected by GBTLs)
are quite similar, we will discuss them collectively as interface
diffusion. The diffusion calculations in PRD are based on the transition
events detected by the method, which involve only part of the atoms
located in the chosen probe region. Considering only the diffusing
atoms (i.e., those which participate in diffusion events) allows us
to identify the diffusion mechanisms and rank their rates. However,
this approach significantly overestimates the diffusion coefficients
because only the most frequent events are counted. Considering all
atoms in the probe region provides more accurate diffusion coefficients
but rescales the contributions of the individual diffusion mechanisms
to the overall diffusivity according to fractions of the respective
defects in the chosen region. Both approaches were considered in this
work.

Including only the diffusing atoms (Fig.~\ref{fig:PRD diffusion}b,d),
the diffusivity decreases in the order (interface and lattice Al by
the vacancy mechanism) $>$ (interface Al and Si by the interstitial
mechanism) $>$ (lattice Al and Si by the interstitial mechanism).
For brevity, by the interstitial mechanism we mean both the direct
and the interstitialcy mechanisms. The results suggest that diffusion
by Al vacancies is the fastest process. Furthermore, as predicted
by the NEB calculations, both Al and Si atoms diffuse faster in the
interface region than in the lattice, although the difference is less
than an order of magnitude. The diffusivities of Al and Si by the
interstitial/interstitialcy mechanism are close to each other, which
is also consistent with the NEB calculations showing that the Si and
Al diffusion barriers of interstitial/interstitialcy events are similar. 

Considering all atoms in the interface and lattice regions, all diffusion
coefficients become significantly smaller than in the previous case,
but they could still be measured by PRD (Fig.~\ref{fig:PRD diffusion}a,c).
The diffusivity decreases in the order (interface Si by the interstitial
mechanism) $>$ (lattice Si by the interstitial mechanism) $>$ (interface
Al by the interstitial mechanism) $>$ (lattice and interface Al by
the vacancy mechanism) $>$ (lattice Al by the interstitial mechanism).
As above, interface diffusion is faster than lattice diffusion. However,
the Si diffusion coefficients are now larger than those of Al diffusion.
The effect is primarily due to the larger number of interstitials
in the deposited Al(Si)/Si layer compared to vacancies, with the difference
reaching a factor of 20 to 30 (see Supplementary Information for details
of the point-defect computations). Another factor is that the probe
region now contains more non-diffusing Al atoms than Si atoms, which
further reduces the measured Al diffusivity. Note that there is no
significant diffusion anisotropy of Al or Si atoms in the interface
region or the lattice. 

\section{Conclusions}

We have performed an MD simulation study of Al and Si diffusion in
${\text{Al(Si)}\{111\}}\parallel{\text{Si}}\{111\}$ IPBs created
by simulated vapor deposition. Besides the IPBs, the deposited Al(Si)
layers contain other structural elements, including GBs, GBTLs, GBTJs,
TDs, and disconnections. The diffusivity in all these structures has
been computed for comparison with IPBs. We find that the IPB diffusivity
is too low to be measured by conventional MD simulations. Previous
work \citep{Al-Si-IPB,Chesser:2024aa} arrived at a similar conclusion.
Therefore, we have applied the PRD method to investigate the diffusion
mechanisms and quantify the IPB diffusion coefficients. The main conclusions
of this work can be summarized as follows: 
\begin{enumerate}
\item The Si surface reconstruction prior to the Al(Si) deposition affects
the obtained IPBs. The Si(111)-1$\times$1 and Si(111)-7$\times$7
surface reconstructions give rise to different orientation relationships
and different IPB structures.
\item It was previously suggested \citep{wu2022atomistic} that Al/Si (7$\times$7)
IPBs could be semi-coherent considering the significant lattice mismatch.
Our simulations indicate that both the (1$\times$1) and (7$\times$7)
IPBs have uniform incoherent structures. We observed no misfit dislocations
at any chemical composition or temperature. 
\item At all temperatures studied in this work, Si diffusion in the substrate
is virtually frozen out. A few Al atoms penetrate into the Si substrate
during the deposition process, but their concentration is too small
to evaluate the diffusivity. In contrast, the concentration of Si
atoms in the deposited Al(Si) layer is higher and their diffusivity
could be measured. We find that Si exhibits a higher diffusivity in
Al(Si) than Al. 
\item The Al and Si diffusion coefficients in the (1$\times$1) and (7$\times$7)
IPBs are smaller than $10^{-13}\,\text{m}^{2}/\text{s}$ at all temperatures
studied here. They are only slightly higher than the lattice diffusion
coefficients in Al(Si), with the difference being less than an order
of magnitude. The IPB diffusion is dominated by the same interstitial
and vacancy diffusion mechanisms as Al(Si) lattice diffusion. GBTLs
in the (1$\times$1) IPB can significantly accelerate the interface
diffusion. No significant diffusion anisotropy was found in the IPBs. 
\item All three types of GBs studied in this work ($\Sigma$3, $\Sigma$7,
and $\Sigma$21) exhibit higher diffusion coefficients than the ${\text{Al}\{111\}}\parallel{\text{Si}}\{111\}$
IPBs, with the difference being more than one order of magnitude.
Si diffusion in the GBs is faster than Al diffusion. No significant
diffusion anisotropy has been observed in the GBs. 
\item The GBs form GBTLs when they intersect with IPBs. The GBTL diffusion
closely correlates with the respective GB diffusion. If the GB has
a higher diffusivity, so does the corresponding GBTL. Overall, GBTL
diffusion is slower than the GB diffusion. GBTJ diffusion is close
to that of the GB with the highest diffusivity to which it is connected.
Among the line defects, GBTJs exhibit a higher diffusivity than GBTLs
and threading dislocations. 
\end{enumerate}
\bigskip{}

\noindent \textbf{Acknowledgements}

\noindent This research was supported by the U.S.~Department of Energy,
Office of Basic Energy Sciences, Division of Materials Sciences and
Engineering, under Award \# DE-SC0023102.

%\bibliographystyle{/Users/ymishin/YURI/Bibliography/ActaMatnew}
%\bibliography{/Users/ymishin/YURI/Bibliography/literat}

\newpage{}

\newpage{}

\clearpage{}
\noindent \begin{center}
{\LARGE{}SUPPLEMENTARY INFORMATION}{\LARGE\par}
\par\end{center}

\section{Si segregation at grain boundaries}

Si atoms strongly segregate to the grain boundaries (GBs) in Al(Si)
deposited on the Si(111)-1$\times$1 substrate. We computed the average
Si concentration in different types of GBs in Al(Si) layers deposited
at different temperatures. The results are shown in Fig.~\ref{fig: Si segregation}.
The calculation was as follows \citep{Koju:2020ab}. We separated
the GBs into 3 nm long segments in the interface plane. Each segment
was 2 nm thick and was centered at the GB core. Then, we computed
the number of Si and Al atoms in each segment. The amount of Si segregation
$[N_{Si}]$ was quantified by the excess number of Si atoms per unit
area $A$ relative to the average Si concentration $c$ in Al(Si):
\[
[N_{Si}]=\frac{(N_{Si}-Nc)}{A}.
\]
Here, $N$ and $N_{Si}$ denote the total number of atoms and the
number of Si atoms within the GB segment, respectively. The results
were averaged over all GB segments of the same type. The error bars
in Fig.~\ref{fig: Si segregation} represent the standard deviation
from the average. As can be seen from Fig.~\ref{fig: Si segregation},
overall, $\Sigma$7 boundaries exhibit the highest Si segregation,
while $\Sigma$3 boundaries the lowest. This order correlates with
the GB diffusivities reported in the main text, where the $\Sigma$7
boundaries exhibit the highest diffusivity and $\Sigma$3 boundaries
the lowest. With increasing temperature, Si segregation at all three
types of GBs increases, while the rate of change with temperature
decreases in the order $\Sigma$7 $>$ $\Sigma$21 $>$ $\Sigma$3.

% Restart figure numbering in appendix from S1
\global\long\def\thefigure{S\arabic{figure}}%
 \setcounter{figure}{0}

\begin{figure}[h]
\centering \includegraphics[width=0.5\linewidth]{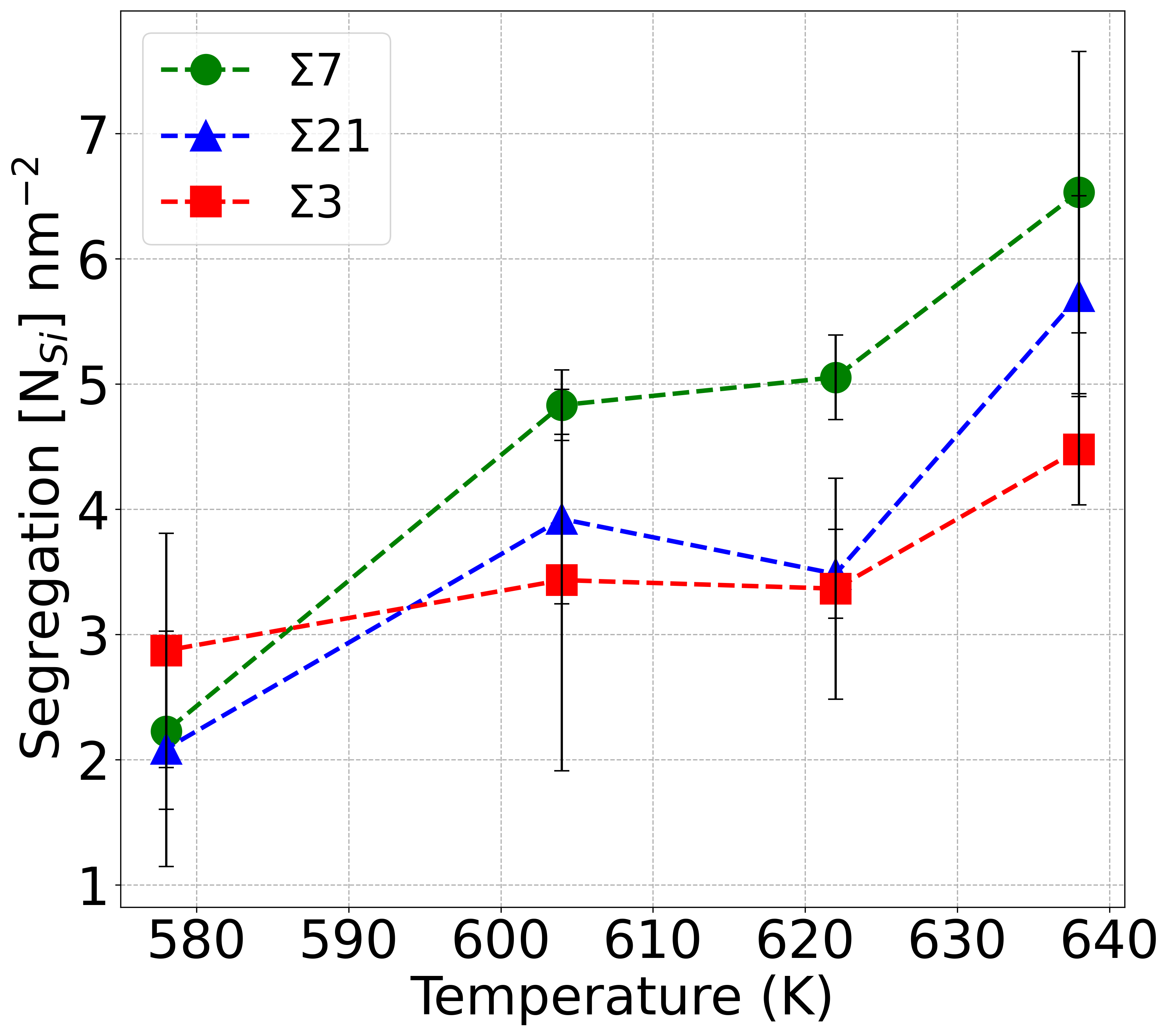}
\caption{Si segregation at $\Sigma$3, $\Sigma$7, and $\Sigma$21 grain boundaries
in the Al(Si) layers deposited on the Si(111)-1$\times$1 substrates
at temperatures ranging from 578 K to 638 K.}
\label{fig: Si segregation}
\end{figure}

\section{Calculation of diffusion coefficient using parallel-replica dynamic
method}

The parallel-replica dynamics (PRD) method \citep{Voter98,Perez:2015aa,Perez:2009aa}
is applied to investigate Al and Si diffusion in IPBs and other structures
of the Al-Si system. This method accelerates MD simulations by exploiting
parallel computing. The basic idea is to run multiple replicas of
the system simultaneously on different processors to sample rare events
and longer timescales. 

The procedure involves the following steps: 
\begin{enumerate}
\item The simulation system is replicated $M$ times, with each replica
assigned to a different processor. 
\item Each replica runs an independent MD simulation starting from the same
initial state. 
\item When any replica detects a transition, all replicas stop the simulations. 
\item The system state after the transition is replicated and sent to all
processors. 
\item All replicas restart their simulations from this new state, and the
process repeats. 
\end{enumerate}
By this algorithm, the PRD effectively extends the timescale of MD
simulations by up to a factor of $M$. 

In this work, we used $M=48$ or $96$ replicas, depending on the
transition frequency in different structures. The MD integration time
step was set to 0.0005 ps. A state transition was defined as an event
in which any atom moves more than 2.5 \AA\ from its previous position
upon energy minimization. The minimization and checks of transitions
were performed every 0.05 ps. Each replica performed a dephasing step
of 0.2 ps upon entering a new state to randomize the initial conditions
and eliminate correlations between replicas. A correlation time of
0.25 ps was set to define the correlation between two consecutive
events. The temperature of PRD simulations is controlled with the
NVT ensemble. 

The diffusion coefficients were calculated from the Einstein relations:
\begin{align}
D_{x} & =\frac{\langle x^{2}\rangle\rho_{\text{deposition}}}{2t\rho_{\text{PRD}}}=\frac{1}{2t}\frac{\rho_{\text{deposition}}}{\rho_{\text{PRD}}}\frac{1}{N}\sum_{i=1}^{N_{\text{event}}}\sum_{j=1}^{N}\left|\Delta x_{ij},\text{if }\left|\Delta r_{ij}\right|>1\right|^{2},\label{eq1}\\
D_{y} & =\frac{\langle y^{2}\rangle\rho_{\text{deposition}}}{2t\rho_{\text{PRD}}}=\frac{1}{2t}\frac{\rho_{\text{deposition}}}{\rho_{\text{PRD}}}\frac{1}{N}\sum_{i=1}^{N_{\text{event}}}\sum_{j=1}^{N}\left|\Delta y_{ij},\text{if }\left|\Delta r_{ij}\right|>1\right|^{2},\label{eq2}\\
D_{x} & =\frac{\langle x^{2}\rangle}{2t}=\frac{N_{\text{event}}}{2tN_{\text{diffuse}}}\sum_{i=1}^{N_{\text{event}}}\sum_{j=1}^{N}\left|\Delta x_{ij},\text{if }\left|\Delta r_{ij}\right|>1\right|^{2},\label{eq3}\\
D_{y} & =\frac{\langle y^{2}\rangle}{2t}=\frac{N_{\text{event}}}{2tN_{\text{diffuse}}}\sum_{i=1}^{N_{\text{event}}}\sum_{j=1}^{N}\left|\Delta y_{ij},\text{if }\left|\Delta r_{ij}\right|>1\right|^{2},\label{eq4}
\end{align}
where $t$ is the aggregate time across all replicas. Eqs.~(\ref{eq1})
and (\ref{eq2}) calculate the diffusion coefficients of all atoms
in the probe region along the $X$ and $Y$ directions. Eqs.~(\ref{eq3})
and (\ref{eq4}) calculate the diffusion coefficients of only the
diffusing atoms in the same probe region. 

In Eqs.~(\ref{eq1}) and (\ref{eq2}), $\rho_{\text{deposition}}$
and $\rho_{\text{PRD}}$ denote the density of interstitials or vacancies
in the system obtained by the simulated deposition process and in
the respective small model created for the PRD simulations. $N$ is
the total number of Al or Si atoms in the probe region. Thus, Eqs.~(\ref{eq1})
and (\ref{eq2}) rescale the MSD by the ratio of (i.e., $\rho_{\text{deposition}}/\rho_{\text{PRD}}$).
In Eqs.~(\ref{eq3}) and (\ref{eq4}), $N_{\text{diffuse}}$ is the
accumulated number of atoms that participate in the diffusive events
(i.e., the total number of atoms that have displacements larger than
1 \AA\ in all the events), and $N_{\text{event}}$ is the total number
of diffusive events in the PRD simulation. Thus, the term $N_{\text{diffuse}}/N_{\text{event}}$
represents the average number of diffusing atoms per diffusion event.
Further, $\Delta x_{ij}$ and $\Delta y_{ij}$ are the displacements
of the $j^{\text{th}}$ atom in the $i^{\text{th}}$ diffusion event
along the $X$ and $Y$ directions, respectively, and $\left|\Delta r_{ij}\right|$
is the displacement magnitude of the $j^{\text{th}}$ atom in the
$i^{\text{th}}$ diffusion event. 

Only atoms that move more than 1 \AA\ in the diffusion event are considered
in the calculation. Because the atoms that displace less than 1 \AA\ return
to their previous positions in the subsequent event, they are assumed
to be vibrating and not diffusing. This criterion works for most conditions
and only fails in very rare cases, with minimal impact on the final
calculation of the diffusion coefficient. 

PRD simulations show that interstitials and vacancies are the major
sources of diffusion in the Al(Si)/Si systems with (7$\times$7) IPBs,
which contain no other microstructure elements. Without interstitials
or vacancies, few diffusion events were observed. Although the formation
of Frenkel pairs can occur, the vacancy and interstitial always recombine
in the subsequent event, contributing nothing to diffusion. The PRD
models have much higher interstitial and vacancy densities than the
deposited structures. The densities of interstitials and vacancies
in the deposited structures were computed using Wigner-Seitz defect
analysis in OVITO \citep{stukowski2012structure}. The interstitial
density in the deposited structures was found to be about 1 interstitial
per 4000 atoms, while that of vacancies is about 1 vacancy per $10^{5}$
atoms (i.e., $\rho_{\text{deposition}}$). The interstitial and vacancy
density in the PRD models is around 1 interstitial or vacancy per
300 to 500 atoms, depending on specific models (i.e., $\rho_{\text{PRD}}$). 

For each PRD model, we ran three PRD simulations starting from different
initial states (i.e., different initial positions of the interstitial
or vacancy in the model). We also ensured that more than 100 diffusion
events occurred in each PRD simulation. The final diffusion coefficient
was the average of three PRD simulations, and the error was estimated
by their standard deviation.
\end{document}